\documentclass[aps,prb,superscriptaddress,onecolumn]{revtex4-2}

\usepackage{amsmath,bm}
\usepackage{amssymb}
\usepackage{graphicx}
\usepackage{subcaption}

\usepackage{slashed}
\usepackage{amsfonts}
\usepackage{verbatim}
\usepackage{appendix}
\usepackage{xcolor}

\usepackage{multirow}
\usepackage{mathrsfs}
\usepackage[colorlinks, breaklinks=true,linkcolor=red, citecolor=blue, linktocpage=true]{hyperref}
\usepackage{footmisc}

\def\bt{\begin{equation}}
\def\bea{\begin{eqnarray}}
\def\ee{\end{equation}}
\def\eea{\end{eqnarray}}

\makeatletter
\begin{document}

	\title{Magnetic Response of a  Two-Dimensional Viscous Electron Fluid}

	\author{Ayd\i n Cem Keser}
	\affiliation{School of Physics, University of New South Wales, Sydney, NSW 2052, Australia}
	\affiliation{Australian Research Council Centre of Excellence in Low-Energy Electronics Technologies,The University of New South Wales, Sydney 2052, Australia}
	
	\author{Oleg P. Sushkov}
	\affiliation{School of Physics, University of New South Wales, Sydney, NSW 2052, Australia}
	\affiliation{Australian Research Council Centre of Excellence in Low-Energy Electronics Technologies,The University of New South Wales, Sydney 2052, Australia}

	\date{\today}

\begin{abstract} 
It has been established that the Coulomb interactions can transform the electron gas into a viscous fluid. This fluid is realized in a number of platforms, including graphene and two-dimensional semiconductor heterostructures. The defining characteristic of the electron fluid  is the formation of layers of charge carriers that are in local thermodynamic equilibrium, as in classical fluids. In the presence of nonuniformities, whirlpools and non-trivial flow profiles are formed, which have been directly imaged in recent experiments.
In this paper, we theoretically study the response of the electron fluid to localized magnetic fields. We find that the electric current is suppressed by viscous vortices in regions where magnetic field is sharply varying, causing strong transport signatures.  Experimentally, our considerations are relevant since local magnetic fields can be applied to the system through implanting adatoms or embedding micro-magnets in the top-gate. Our theory is essential for the characterization and future applications of electron fluids in hydrodynamic spin transport.

\end{abstract}

\maketitle

\section{Introduction}
\label{sec:intro} 
	The hydrodynamic transport of electrons has been the subject of intense research in recent years.~\cite{narozhny_hydrodynamic_2022,keser_geometric_2021,polini_viscous_2020,lucas_hydrodynamics_2018} 
	Electron transport is usually dominated by impurity scattering in ordinary conductors, where the electrons behave like a diffusive gas. In  sufficiently pure samples, Coulomb interactions transform the electron system into a viscous fluid.~\cite{polini_viscous_2020,lucas_hydrodynamics_2018,narozhny_hydrodynamic_2017,zaanen_electrons_2016} The  early  prediction of an `electron fluid' in solids, made  by Gurzhi~\cite{gurzhi_hydrodynamic_1968,gurzhi_minimum_1963}  has been confirmed experimentally with the discovery of high purity samples.~\cite{de_jong_hydrodynamic_1995}  
	
	
Notable consequences of fluid behavior that has been demonstrated by recent studies are: the decrease of resistance with increasing temperature (Gurzhi effect)\cite{de_jong_hydrodynamic_1995,krishna_kumar_superballistic_2017,keser_geometric_2021, ginzburg_superballistic_2021},	Poiseuille flow profiles \cite{sulpizio_visualizing_2019,ku_imaging_2020,braem_scanning_2018,ella_simultaneous_2019},
negative nonlocal resistance and formation of whirlpools \cite{govorov_hydrodynamic_2004,bandurin_negative_2016},
Hall viscosity \cite{gusev_viscous_2018,berdyugin_measuring_2019}, negative magnetoresistance,~\cite{shi_colossal_2014,keser_geometric_2021,mani_size-dependent_2013} the violation of the Wiedemann-Franz
law \cite{crossno_observation_2016,gooth_thermal_2018, jaoui_thermal_2021,lucas_electronic_2018,ahn_mesoscale_2022}, anomalous scaling of resistance with channel width, \cite{gooth_thermal_2018,moll_evidence_2016} and quantum-critical dynamic conductivity.~\cite{gallagher_quantum-critical_2019}

	Moreover, many novel phenomena have been proposed, such as hydrodynamic spin transport~\cite{tatara_hydrodynamic_2021,doornenbal_spinvorticity_2019} inspired by experiments on liquid mercury.~\cite{takahashi_spin_2016,takahashi_giant_2020} 

	The defining characteristic of hydrodynamic transport is the formation of layers of charge carriers that are in local thermodynamic equilibrium, as in classical fluids.~\cite{landau_fluid_2013} In uniform flow, the layers are at rest with respect to a common inertial frame. On the other hand, in the presence of nonuniformities, whirlpools and non-trivial flow profiles are formed. These flow patterns can be directly imaged, as in recent sophisticated experiments.~\cite{sulpizio_visualizing_2019,ku_imaging_2020,braem_scanning_2018,ella_simultaneous_2019}
	
	In this paper, we focus on the magnetic response of the two-dimensional (2D) electron fluid. Our motivation is two-fold. Firstly, magnetic response of classical fluids is a curious phenomenon called `Moses effect'~\cite{bormashenko_moses_2019}  with applications in colloidal chemistry. Akin to this effect, we find that the electron fluid streamlines deform to avoid regions where magnetic fields vary sharply.  

	Secondly, the magnetic manipulation of the 2D electron fluid is now a practical reality in a number of platforms such as Co adatoms on graphene~\cite{ren_kondo_2014,uchoa_localized_2008,wehling_orbitally_2010,brar_gate-controlled_2011}, or micromagnets embedded in the top gate of a semiconductor heterostructures~\cite{neumann_simulation_2015, engdahl_micromagnets_2022}. Moreover, magnetic response of the electron fluid is essential for its characterization and future technological applications, such as hydrodynamic spin transport.~\cite{takahashi_giant_2020}
	
	In an ultrapure system, when the temperature is comparable to the Fermi energy, electrons scatter much more often with each other, than they do with phonons and impurities.~\cite{narozhny_hydrodynamic_2022,keser_geometric_2021}. As a result, local thermodynamic equilibrium is established within a region of size comparable to the electron-electron (e-e) scattering length. Above this length scale, electrons can be described like a classical fluid that obeys the Navier-Stokes equation.~\cite{lucas_hydrodynamics_2018,landau_fluid_2013} In the presence of an out-of-plane magnetic field, the Lorentz force affects the electron fluid. A uniform magnetic field can not deflect the electron flow owing to the incompressibility of electron fluid, and merely creates a Hall pressure (voltage), whereas nonuniform fields can have a non-trivial effect on the flow profile.
	
The main findings of our paper are:  i) that the nonuniform localized magnetic field creates vorticity in the system, ii) a magnetic impurity of size comparable to the e-e scattering length induces Lorentz force gradients so strongly that, current flow is suppressed near the impurity and electron flow is expelled out of this vicinity, and finally iii) the viscous energy loss due to magnetically induced vortices cause large resistive contributions to transport.

The plan of the paper is as follows. 
In Sec.~\ref{sec:magnetohydro} we  write down the fluid equations of motion that describes the system. 
We show that under realistic experimental conditions, the equations of motion become linear in current density, and there is a natural length scale in the system arising from the interaction of electrons with each other and also with phonons and impurities. 
This length scale separates the viscous dominated (ultraviolet) and drag dominated (infrared) regions. 
In Sec.~\ref{sec:stokes} we consider the `magnetic Stokes problem', where a uniform flow is incident on a region that contains out-of-plane magnetic field. 
In Sec.~\ref{sec:soft} we perturbatively determine the flow profile of electrons, when both the magnetic field and its gradients are weak. 
In Sec.~\ref{sec:log}, we show that the perturbative theory breaks down near a magnetic impurity, and a logarithmic renormalization of current density takes effect. In Sec.~\ref{sec:resistance} we calculate the transport signature of localized magnetic fields on the system. Finally, Sec.~\ref{sec:conc} is devoted to the conclusions.


	
	\begin{figure}[th!]
		\centering
		\begin{subfigure}[b]{0.35\textwidth}
			\centering
			\includegraphics[width=\textwidth]{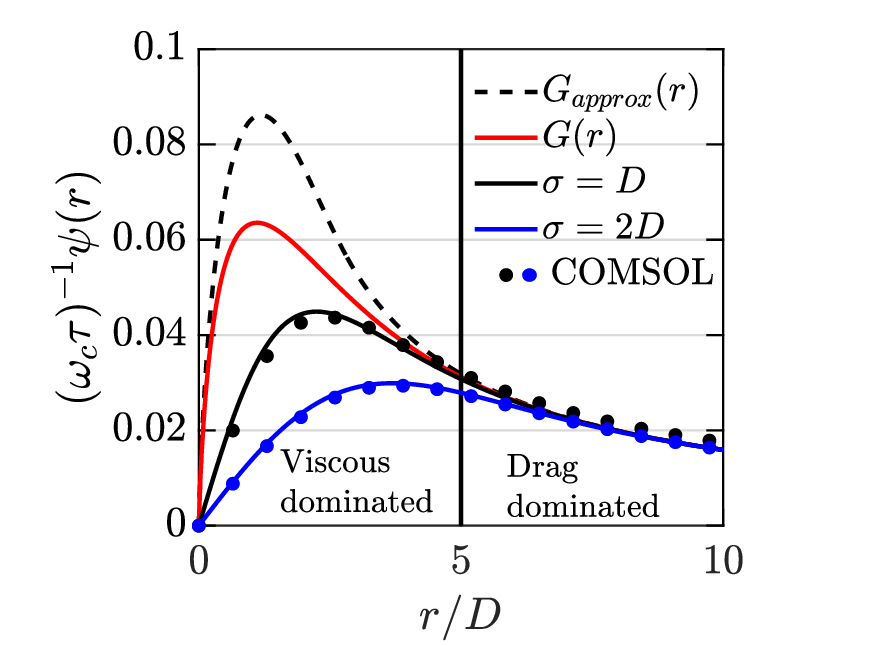}
			\caption{}
			\label{fig:green}
		\end{subfigure}
		\hspace{-1cm}
		\begin{subfigure}[b]{0.35\textwidth}
			\centering
			\includegraphics[width=\textwidth]{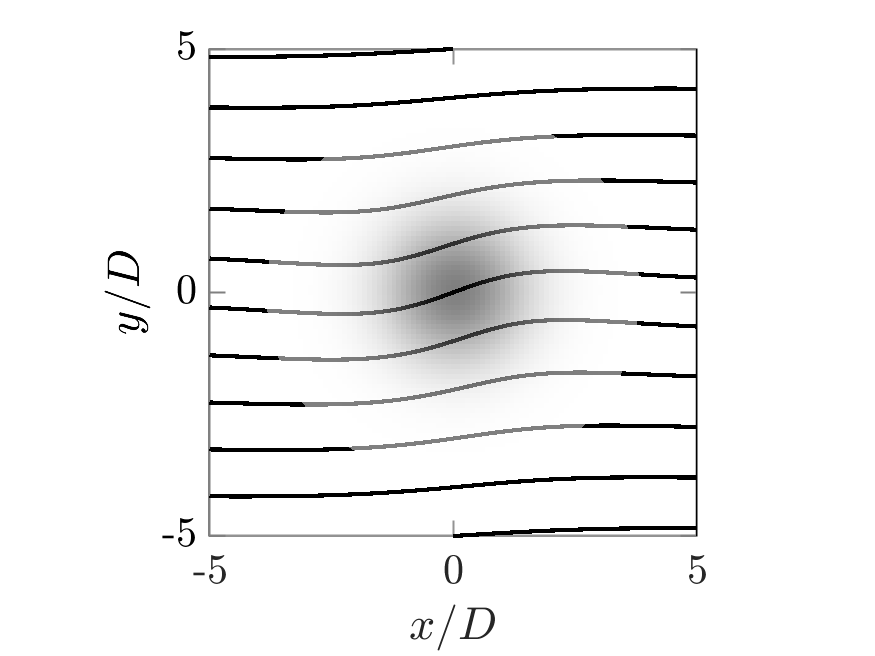}
			\caption{}
			\label{fig:gaussian}
		\end{subfigure}
		\hspace{-1cm}
		\begin{subfigure}[b]{0.35\textwidth}
			\centering
			\includegraphics[width=\textwidth]{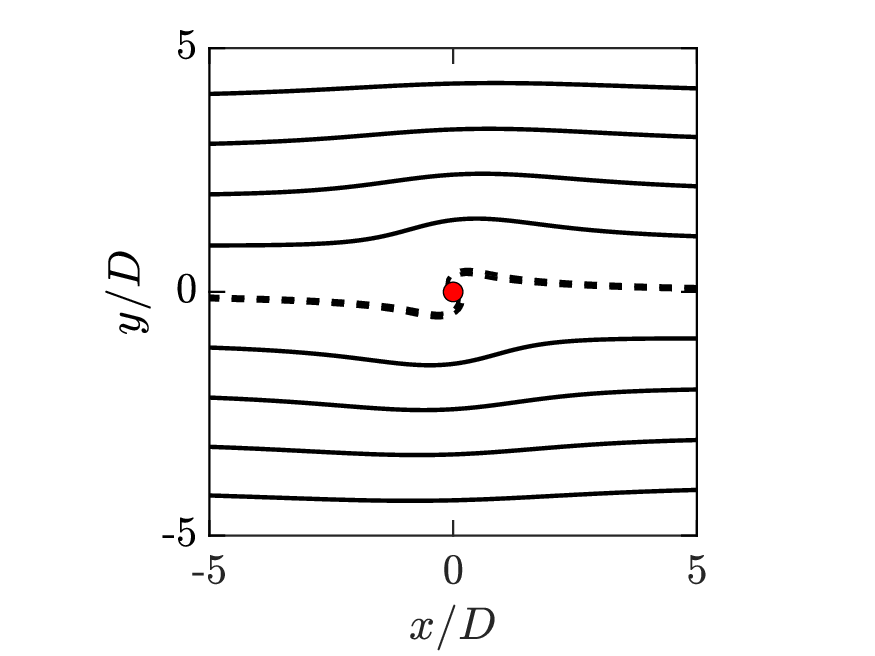}
			\caption{}
			\label{fig:impurity}
		\end{subfigure}
		\caption{The electrons follow curves of constant streamfunction $\Psi$, called the streamlines. In the Stokes problem, we assume a uniform flow of electrons in the $x$-direction is incident on a localized magnetic field region. a) The radial part of the normalized correction to the stream function in the weak field intensity and soft gradient regime. The total streamfunction is $\Psi =  -v_0 y +v_0 D \psi(r)\cos(\phi)$. The out-of-plane magnetic field $\mathbf{B} =\boldsymbol{\hat{z}} B b(r)  $ is assumed to be radially symmetric with a Gaussian profile $b(r,\sigma)$ centered around the origin with a standard deviation $\sigma$. The solid blue and black curves are the analytic solutions (see Eq.~\eqref{psi_k}) for, $\sigma = D, 2D$  respectively. The filled circles are data points obtained by simulating the system on COMSOL Multiphysics\textregistered\: software\cite{comsol_ab_comsol_2015} with weak field parameters.  Notice that the blue circles have a closer agreement with theory, since the analytic solution is obtained in the weak gradient limit.  In the limit $\sigma\to 0$, the profile becomes a delta function, hence we recover the radial part of the Green's function in Eq.~\eqref{Gexact}, shown by solid red curve. The flow is viscous dominated close to the origin and drag dominated further away. The transition occurs around $r=5D$. The Green's function can be piecewise constructed in these asymptotic regions, respectively,  as $G_<,G_>$ in Eq.~\eqref{Gs}, shown by the dashed black curve b) The streamlines of an electric current incident on a magnetic field region with soft Gaussian profile. The current in the $x$-direction is deflected by the out-of-plane magnetic field concentrated in the dark region. c) A magnetic impurity creates a sharply changing magnetic field profile and induces large Lorentz force gradients. Hence, the perturbative solution no longer holds. Instead, the current near the impurity is logarithmically suppressed, therefore the streamline passing through the origin (dashed lined) does not carry current.  Here, the small radius of the impurity $r_0 = 0.2 D$ serves as the ultraviolet (UV) logarithmic cut-off.  In (b) and (c) the magnetic field is exaggerated to make the streamline deformation more prominently visible.
		}
		\label{fig:streams}
	\end{figure}

\section{The Magnetohydrodynamics of the Electron Fluid}
\label{sec:magnetohydro}

In this section we describe the physical model of the 2D electron flow under simultaneous application of electrochemical potential gradients and out-of-plane magnetic field, in terms of the modified incompressible Navier-Stokes equation
\begin{equation}
\mathbf{v}/\tau + \mathbf{v}\cdot \nabla\mathbf{v}- \nu \nabla^2\mathbf{v} = -\nabla\Phi/m^* -|e|/m^* \mathbf{v}\times\mathbf{B}.
\end{equation} 
Here $\mathbf{v}$ is the electron fluid velocity vector field that is incompressible $\nabla\cdot \mathbf{v} = 0$. The velocity field is related to the electric current density and electron  number density as $\mathbf{J} = -|e| n\mathbf{v}$. The collision time $\tau$ is responsible for momentum loss scattering processes such as those due to phonons and impurities, and creates an effective drag on the fluid. The kinematic viscosity $\nu$ is the intrinsic property of the electron fluid calculated from its interaction Hamiltonian. In  2D $\nu = v_F l_{ee}/4$ holds, where $v_F$ is the Fermi velocity and $l_{ee}$ is the e-e scattering length. $\Phi$ is the electrochemical potential, $m^*$ is the effective mass of the electron. Finally, $\mathbf{B}$ is the out of plane magnetic field.   Based on real experiments, for example ~\cite{keser_geometric_2021}, we consider low currents  $\sim 200\:\mathrm{nA}$. with typical velocities about $200\:\text{m}/\text{s}$, in a system with length $~10\:\mu\mathrm{m}$. Meanwhile, the kinematic viscosity is very large, $\nu \sim 200\:\mathrm{cm}^2/\mathrm{s}$ making the electron fluid nearly $300$ times thicker than honey. As a result, the Reynolds number turns out to be $\mathrm{Re}\sim 0.01$. This means then nonlinear Navier term, $\mathbf{v}\cdot \nabla \mathbf{v}$, responsible for inertial forces due to convective acceleration, is a hundred times smaller than viscous forces.   Therefore, we neglect this term and work with the linear Stokes equation. The realistic range of parameters are given in Table~\ref{tab:params}.

It is known that the viscosity of electron fluid depends on the applied magnetic field~\cite{alekseev_negative_2016}, and additional Hall viscous friction emerges~\cite{berdyugin_measuring_2019}. However, both of these effects are second order in the $B$-field and therefore negligible, since  the deviation due to B-field is $\delta\nu/\nu\sim 1\%$.

We can parametrize the Stokes equation as
\begin{equation}
\label{stokes}
\mathbf{v} -D^2 \nabla^2 \mathbf{v}  = -\tau \nabla \Phi/m^* - \omega_c \tau b(\mathbf{x}) \mathbf{v}\times \boldsymbol{\hat{z}},
\end{equation}
where $D=\sqrt{\nu\tau} $ is the momentum diffusion length, which can also be expressed in terms of the geometric average $D = \sqrt{l_{ee}l_{mfp}}/2$ of two length scales. These are, namely, the e-e scattering length $l_{ee}$ that conserves momentum in the electron system  and mean free path of collisions of electrons with impurities and phonons, that subtracts momentum from the electron system. The product   $\omega_c = |e|B/m^*$ is the cyclotron frequency and  $b$ captures the spatial dependence of the magnetic field.
We define the stream function as
\begin{align}
\label{stream}
\nonumber v_y  = \partial_x \Psi,\quad v_x = -\partial_y \Psi,\\
v_r = -\frac{1}{r}\partial_\phi \Psi,\quad v_\phi = \partial_r \Psi,
\end{align}
so that the incompressibility constraint $\nabla \cdot  \mathbf{v} =0$ is automatically satisfied. The curves with constant $\Psi$ values are called streamlines. The velocity field is, by definition, everywhere tangent to the curves of equal $\Psi$, that is, the particles follow the streamlines.

\section{The Magnetic Stokes Problem}
\label{sec:stokes}

We imagine the realistic scenario where a uniform flow of electrons in the $x$-direction is  incident on a magnetic field distribution centered around the origin. The stream function can be decomposed into a part that generates the uniform flow and a correction term due to the magnetic field induced vorticity:
\begin{equation}
\Psi = -v_0 y + v_0 D\psi, \quad \omega = \hat{z}\cdot \nabla \times \mathbf{v} = v_0 D\nabla^2 \psi.
\end{equation}
Now taking the curl of the Stokes equation Eq.~\eqref{stokes} we obtain 
\begin{equation}
\label{magstokes}
\nabla^2 \Psi - D^2\nabla^4\Psi = \omega_c \tau \nabla b(\mathbf{x})\cdot \mathbf{v}.
\end{equation} 
We note that, when the magnetic field is uniform, the right-hand side vanishes and therefore the streamlines are trivial, $\Psi = -v_0 y$. The incoming flow is not affected but a classical Hall voltage, $V_y =- m^* \omega_c v_0 W$, develops in the system with width $W$.

To find the response to a nonuniform field, it is beneficial to find the Green's function that solves the above equation for $b=\delta(x)$ up to normalization constants. Below, we construct the Green's function in weak and strong magnetic fields.

\section{Perturbative Solution in Weak Magnetic field with a Soft Gradient}
\label{sec:soft}
When the magnetic field strength and the gradient is weak, the source term in Eq.~\eqref{magstokes}, $\omega_c \tau \nabla b\cdot \mathbf{v} \ll v_0/D$. Therefore, the corrections to the incoming velocity $v_0$ is small. We therefore apply first order  perturbation theory and take the velocity in the source term on the right-hand side of Eq.~\eqref{magstokes} to zeroth order. 
We can work in the normalized units where
\begin{equation}
   v_0 = D = 1.
\end{equation}
Then the Green's function obeys
\begin{equation}
\label{Green}
\nabla^2 G - \nabla^4 G = \partial_x \delta(\mathbf{x}).
\end{equation}
Before solving this problem exactly, we use scaling analysis to have an intuitive grasp of the problem. In the regions $r> R\sim 1$ and $r<R\sim 1$, the dominant operators are $\nabla^2$ and, $\nabla^4$  respectively.  In the large $r$ regime we may write
\begin{equation}
\nabla^2 G_> = \partial_x \delta(\mathbf{x}).
\end{equation}
We can recognize this from electrostatics and solve it as
\begin{equation}
G_>= G(r\to\infty)\to\frac{1}{2\pi}\partial_x \log(r) = \frac{\cos(\phi)}{2\pi r}.
\end{equation}
Meanwhile for $r<R\lesssim 1$ we have
\begin{equation}
-\nabla^2\omega_< = \partial_x \delta(\mathbf{x}).
\end{equation}
Needless to say this has the same solution, however we can add a function harmonic in the unit circle to the solution:
\begin{equation}
\label{vort_small}
\omega_< = -\frac{\cos(\phi)}{2\pi r} + \frac{R^{-2}}{2\pi} r \cos(\phi).
\end{equation}
This choice is judicious so that $\omega_<(R) = \omega_>(R)=0$ at the matching point. 
Now  solving  for the stream function in polar coordinates, we obtain
\begin{equation}
\partial_r (r^{-1}	\partial_r(r G_<))  = \left(-\frac{1}{2\pi r} + \frac{R^{-2}}{2\pi} r \right)\cos(\phi),
\end{equation}
which implies
\begin{equation}
\label{G_small}
G_<= G(r\to 0) \to \frac{\cos(\phi)}{4\pi}\left(-r\log(r) + Ar + R^{-2}\frac{r^3}{4}\right).
\end{equation}
Note that this solution diverges to infinity if $r\sim R\to\infty$. This infrared (IR) divergence is essentially the Stokes paradox where, the creeping flow solution does not exist in 2D~\cite{landau_fluid_2013}. IR divergence is automatically cured in our problem because momentum loss terms define a finite length scale $D=\sqrt{\nu\tau}\equiv 1$  and therefore, the $G_<$ is only valid in the bounded region $r<R\sim 1$.  

At the boundary of viscous to ohmic flow, $r = R$, we demand $G$ and $\partial_r G$ to be continuous. From these constraints we obtain
\begin{subequations}
	\label{Gs}
	\begin{align}
	G(r\gg R=2\sqrt{2}) &\to \frac{\cos(\phi)}{2\pi r},\\
	G(r\ll R=2\sqrt{2}) &\to\left(\frac{r^3}{64} - \frac{r}{4}\log\left(\frac{r^2}{8}\right)\right)\frac{\cos(\phi)}{2\pi},
	\end{align}
\end{subequations}
the radial part of which is plotted in Fig.~\ref{fig:green} as dashed lines.
The exact Green's function defined as the solution of Eq.~\ref{Green} can be obtained in momentum space as
\begin{equation}
G(k) = -\frac{i k_x}{k^2(1+ k^2)}.
\end{equation}
In real space this is 
\begin{equation}
\label{Gexact}
G = \frac{\cos(\phi)}{2\pi}\int_0^\infty dk  \frac{J_1(kr) }{1+ k^2} = \frac{\cos(\phi)}{2\pi r}[1- rK_1(r)],
\end{equation}
where $K_1$ is the modified Bessel function of the second kind of order 1. Therefore, asymptotic behavior of $G$ matches Eq.~\eqref{Gs}. The radial part of the exact Green's function is shown in Fig.~\ref{fig:green}.
	
Now we can write down the solution for a weak and arbitrary distribution $b(\mathbf{x})$ with soft gradients
\begin{equation}
\label{psi_k}
\psi(\mathbf{k}) = -\omega_c \tau  \frac{i k_x b(\mathbf{k})}{k^2(1+k^2)},
\end{equation}	
where $\psi = \Psi +  y$ is the correction to the stream function in real space, and $\psi(\mathbf{k})$ is its Fourier pair. For a radially symmetric and Gaussian magnetic field distribution $b(r,\sigma)$ with standard deviation $\sigma$, we can plot the radial part of the normalized correction to the stream function in Fig.~\ref{fig:green} for $\sigma = D,2D$. As expected, when $\sigma\to 0$, the Gaussian profile approaches a delta function hence we recover the exact Green's function Eq.~\eqref{Gexact}, up to normalizing constants.

As seen in  Fig.~\ref{fig:gaussian}, the electron trajectories are bent due to Lorentz force gradients. As a measure of this, we can calculate the transverse velocity $v_y(0)$ at the center of the magnetic domain, with a gaussian profile, as
\begin{equation}
v_y(0) = \partial_x \psi = \frac{\omega_c\tau}{8\pi}\int_0^\infty d\alpha \frac{ e^{-\alpha/2}}{\sigma^2+\alpha} = \frac{\omega_c\tau}{8\pi} e^{\sigma^2/2} \text{E}_1(\sigma^2/2),\quad \mathrm{(perturbative)}.
\end{equation}
Here $\mathrm{E}_1(z) = \int_z^\infty dt \exp(-t)/t$ is the exponential integral. When the standard deviation is large, the Lorentz Force gradients are weak and the transverse velocity decays as $v_y(0) \sim v_0\omega_c\tau D^2/(4\pi\sigma^2)$. In the limit of strong gradients, $\sigma\to 0$, we have $v_y(0) \sim \log(D/\sigma)$ and the perturbation theory breaks down as expected. 

\begin{figure}[th!]
		\centering
  	\hspace{-1cm}
		\begin{subfigure}[b]{0.35\textwidth}
			\centering
			\includegraphics[width=\textwidth]{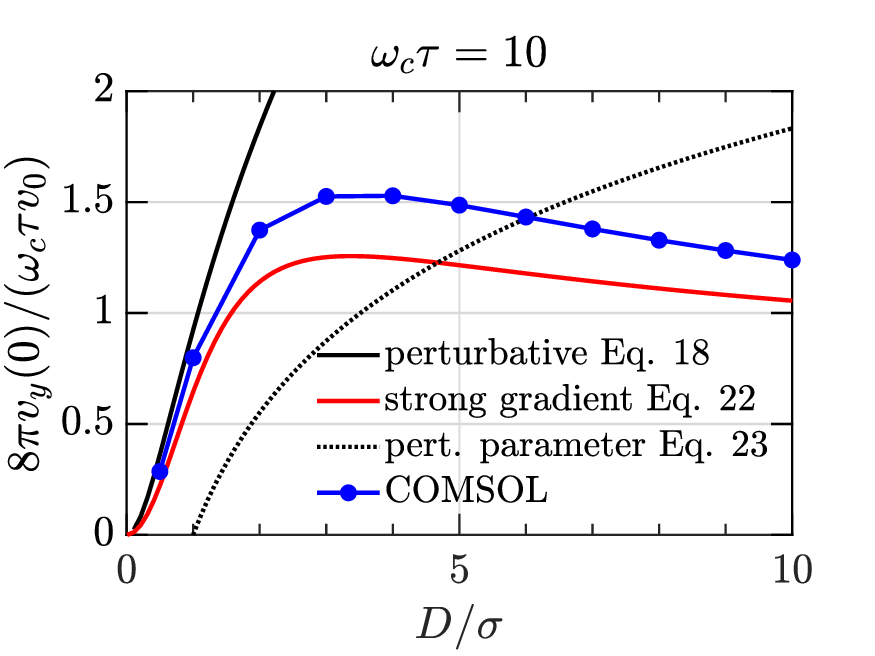}
			\caption{}
			\label{fig:transition}
		\end{subfigure}
		\hspace{0cm}
		\begin{subfigure}[b]{0.3\textwidth}
			\centering
			\includegraphics[width=\textwidth]{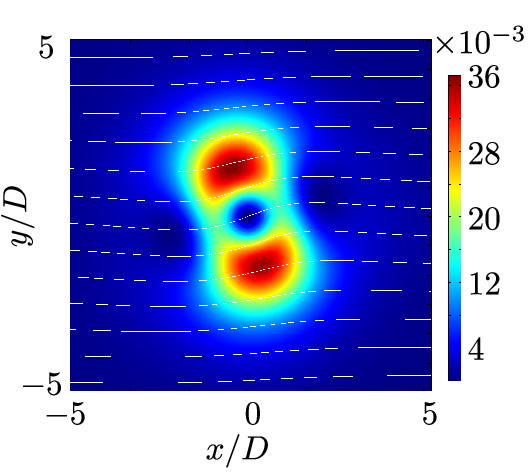}
			\caption{}
			\label{fig:weak_power_density}
		\end{subfigure}
		\hspace{0cm}
		\begin{subfigure}[b]{0.3\textwidth}
			\centering
			\includegraphics[width=\textwidth]{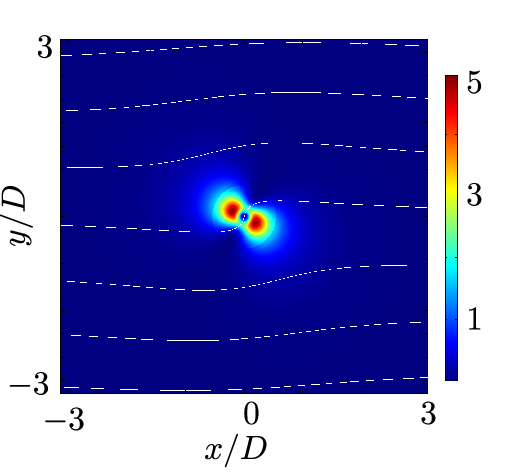}
			\caption{}
			\label{fig:strong_power_density}
		\end{subfigure}
		\caption{a) Normalized transverse velocity at the center of magnetic region, fixed as origin, plotted against the sharpness of the Gaussian magnetic field distribution with standard deviation $\sigma$, which fixes the UV cut-off $r_0=\sigma$. A strong magnetic field $\omega_c\tau =10$ is chosen to highlight the transition from perturbative to strong  gradient regimes. The solid black line shows the soft gradient perturbation theory result, which agrees with the numerical solution, as also indicated in Fig.~\ref{fig:green}. The dotted black line is the perturbation theory parameter $A$ in Eq.~\eqref{pert_param}. The perturbation theory breaks down when the parameter $A\approx 0.5$. The solid red curve is from the strong gradient solution Eq.~\eqref{vxvy}, which shows the logarithmic suppression at the origin. The COMSOL Multiphysics\textregistered\: \cite{comsol_ab_comsol_2015}  simulations, shown by the blue dots, agree with the strong gradient calculation up to logarithmic accuracy.
  b-c) Distribution of viscous power density in real space, normalized to the uniform ohmic dissipation $(n e v_0)^2 \rho_\tau \approx 0.05\:\text{W}/\text{m}^2$. The simulation is performed on COMSOL\textregistered. The white curves are the streamlines. The magnetic field is chosen to be large ($\omega_c\tau = 10$) for better visibility of streamline deformation. The viscous dissipation is due to a Gaussian magnetic field distribution with standard deviation $\sigma$.  The perturbative limit with soft gradient $\sigma=D$ is shown in (b) and the strong field, strong gradient $\sigma = D/10$ is realized in (c).   Note that the power dissipation is suppressed at the origin, which is a consequence of the Gaussian magnetic field distribution.	}
		\label{fig:power}
	\end{figure}

\section{Logarithmic Suppression of Velocity Near a Magnetic Impurity in the Strong Field, Strong Gradient Limit }
\label{sec:log}

A magnetic impurity with size $r_0\ll D$  creates a large magnetic field gradient, that is, the source term in Eq.~\eqref{magstokes} becomes large $\omega_c \tau \nabla b\cdot \mathbf{v} \gg v_0/D$ . Therefore, the conditions for a perturbative reduction of the magnetic Stokes problem Eq.~\eqref{magstokes} to Eq.~\eqref{Green} are not fulfilled. 

Once again, we work in the normalized units where
\begin{equation}
D = v_0 = 1,
\end{equation} 
unless otherwise stated, and reintroduce $D$ or $v_0$, wherever convenient. We restore physical units in the plots and in calculating the estimates. 
Modelling the magnetic impurity as a delta function, we solve Eq.~\eqref{magstokes} as
\begin{equation}
\label{no-pert}
\psi(\mathbf{k}) = -\omega_c \tau\frac{i \mathbf{k}\cdot\mathbf{v}(r\to 0)}{k^2(1+k^2)}.
\end{equation}
where $\psi = \Psi + y$  and $\psi(\mathbf{k})$ is its Fourier pair.  For self-consistency, we require the velocity near the origin to satisfy the relations in Eq.~\eqref{stream}:
\begin{subequations}
	\begin{align}
	v_x(0) &= 1 +  \omega_c \tau \int \frac{d^2 k}{(2\pi)^2} i k_y  \frac{i k_x v_x(0) + i k_y v_y(0)}{k^2(1+k^2)}
	,\\
	v_y(0) &=  -\omega_c \tau\int \frac{d^2 k}{(2\pi)^2} i k_x  \frac{i k_x v_x(0) + i k_y v_y(0)}{k^2(1+k^2)}.
	\end{align}
\end{subequations}
Solving these we have
\begin{equation}
\label{vxvy}
	v_x(0) \to  \frac{1}{1+A^2} ,\quad
	v_y(0) \to\frac{A}{1+A^2},
\end{equation}
where,  after restoring $D$, 
\begin{equation}
\label{pert_param}
	A = \frac{\omega_c \tau}{8\pi}\log(1+D^2/r^2_0)\to \frac{\omega_c \tau}{4\pi}\log(D/r_0),
\end{equation}
and $r_0\ll D$ is the ultraviolet cut-off scale. This scale is determined by the physical size of the impurity and the e-e interaction length $l_{ee}$. Below $r<l_{ee}\sim 200\:\text{nm}$, hydrodynamic theory progressively breaks down as the electrons can not  locally equilibrate to form the fluid. Therefore, realistically, the logarithm can not grow more than $\log(D/l_{ee})\sim 1$. This means that a transition from perturbation theory to strong field regime should occur for large $\omega\tau$ leading to $A\lesssim 1$. Fig.~\ref{fig:transition} shows that the transition from soft to strong gradient regimes, and the logarithmic suppression occurs around $A\approx 0.5$, that is equivalent to $\omega_c\tau \approx 6$ or $B\approx 90\:\text{mT}$ according to Table~\ref{tab:params}. The numerical simulations agree with the strong field theory up to logarithmic accuracy. 

We can find $\psi$ of Eq.~\eqref{no-pert} in real space, by using the result in Eq.~\eqref{Gexact}. When $r\ll D$ we have
\begin{equation}
\Psi  = - y + \frac{\omega_c\tau}{4\pi} (v_x(0) x + v_y(0) y) \log(D/r).
\end{equation}
We note that, the 0th streamline that passes close to the origin $r\sim r_0$ (with value $\psi \sim 1/\log^2(D/r_0)$) does not carry any velocity since, 
\begin{equation}
v_x|_{y\to r_0} = -\partial_y \psi|_{y\to r_0} \to 0,
\end{equation}
meaning that the fluid avoids the singularity by deforming around it. The streamlines are shown in Fig.~\ref{fig:impurity}.
This situation is a similar to the Moses effect, where a diamagnetic fluid evacuates regions of magnetic field.~\cite{bormashenko_moses_2019} Here, the Lorentz force gradient on the charge fluid becomes very large near the impurity and expels the flow lines outwards. 

\begin{table}[th!]
	\caption{Parameters of electron system at $30$-$40\:\text{K}$, based on the experiment in Keser \textit{et al.}~\cite{keser_geometric_2021}. }
	\medskip
\centering\renewcommand{\arraystretch}{1.2}
	\begin{tabular}{lllll}
		\hline\hline
		Parameter & Expression & Value & Description \\
		\hline
		$m^*$	& $0.067 m_e $ & $6.1\times 10^{-32} \:\text{kg} $  & Effective electron mass \\
		$n$	& & $2.45 \times 10^{11} \:\text{cm}^{-2}$      & Electron density   \\
		$E_F$	& $\hbar^2 \pi n/m^*$  & $ 8.8 \:\text{meV}$,  $100 \:\text{K}$  &  Fermi energy\\
		\hline
		$l_{ee}$ & & $ 200 \: \text{nm}$ & e-e interaction length scale\\
		$l_{mfp}$ & $v_F\tau_{mfp}$  & $5.9 \: \mu\text{m} $ & Mean free path \\ 
		$D$ & $\sqrt{l_{ee}l_{mfp}}/2$ & $0.54 \:\mu \text{m}$ & Momentum diffusion length \\
		
		\hline
		$\tau_{mfp}$ &  & $27.7 \: \text{ps}$ & Momentum relaxation time \\
		$\rho_{\tau}$ & $(n e^2 \tau/m^*)^{-1}$ & $35 \: \Omega $ & Ohmic resistivity\\
		$\mu$ & $e\tau_{mfp}/m^*$ & $0.7$  million $\text{cm}^2/(\text{Vs})$ & electron mobility\\
		\hline
		$B^*$ & $m^* v_F/(2 e l_{ee})$ & $204.2\:\text{mT}$ & Characteristic viscous magnetic field\\
		$\phi_c^* = \omega_c^*\tau$& $e B^*\tau/m^*$ &$14.9$ & Cyclotron parameter at characteristic field
	\end{tabular}
	\label{tab:params}
\end{table}

\section{Viscous Dissipation and Resistance}
\label{sec:resistance}

The deformation of streamlines as shown in Fig~\ref{fig:streams} results in viscous stresses. From the Stokes Eq.~\eqref{stokes}, we can find the rate of energy dissipation due to the viscous forces as the integral of the viscous power density (that is the work done against the viscous force per unit time):
\begin{equation}
\dot{E}_{\nu} = -\nu m^* n \int d^2 x\mathbf{v}\cdot \nabla^2 \mathbf{v} ,
\end{equation}
where $n$ is the number density of electrons. Figs.~\ref{fig:weak_power_density} and \ref{fig:strong_power_density} show the distribution of viscous power dissipation in real space for weak and strong gradient limits.  By using the definition Eq.~\eqref{stream} of the stream function, and using Fourier methods we can write the dissipation in physical units, in terms of a dimensionless integral
\begin{equation}
\dot{E}_{\nu} = \nu m^* n v_0^2 D^2 \int \frac{d^2 k}{(2\pi)^2} k^4 |\psi(\mathbf{k})|^2.
\end{equation}
Since  $\dot{E}_{\nu} = R_{\nu} I^2$,  we can find the induced viscous resistance in a system with an effective width $W$, for which the current is $I = W |e| n v$. We express the viscous resistance in terms of the Drude resistance $\rho_\tau = m^*/(n e^2 \tau)$ as 
\begin{equation}
R_\nu/\rho_\tau = \frac{D^2}{W^2} (\omega_c  \tau)^2 f(D/r_0),
\end{equation}
for some dimensionless form factor $f$ that depends on the characteristic length scale of the magnetic field distribution $r_0$.

In a system with length $L$ and width $W$ that contains magnetic impurities with number density $n_{m}$, the contribution to resistivity $\rho_{\nu,m} = R_{\nu} W/L$ is
\begin{equation}
\rho_{\nu,m} =  \rho_\tau D^2 n_{m}  (\omega_c  \tau)^2 f.
\end{equation} 
For a radially symmetric soft magnetic field distribution $b(r,\sigma)$ considered in Sec.~\ref{sec:soft}, this form factor is
\begin{equation}
\label{fana}
f(s = \sigma/D) =  \int_0^\infty \frac{d \alpha}{8\pi}  \frac{ \alpha e^{-s^2 \alpha} }{(1+\alpha)^2}= \frac{-1+(1+s^2) e^{s^2}\text{E}_1(s^2)}{8\pi},\quad \text{(perturbative),}
\end{equation}
where $\text{E}_1$ is the exponential integral. The exponential suppression of the integrand in $k$ space is reflected in Fig.~\ref{fig:weak_power_density} and \ref{fig:strong_power_density} as suppression of power density at the origin.
When $\sigma \gtrsim D$, the denominator of the integrand is $(1+\alpha)^2 \sim 1$, within the relevant range of integration. Therefore, for soft magnetic field profile, the function $f$ decays rapidly as $f\sim D^4(8\pi\sigma^4)^{-1}$.
The hydrodynamic theory becomes less accurate at scales smaller than the e-e interaction length $l_{ee}$, at which thermodynamic equilibration is established. For this reason, we shall take $r_0 \sim l_{ee}$ and therefore  $f\approx 1/(8\pi)$. 

In the delta function limit $\sigma/D \to r_0/D \ll 1$, the term $e^{-s^2\alpha}\to 1$,  in the numerator of the integrand, hence we have a logarithmic divergence within the perturbative approach:
\begin{equation}
f\to \frac{1}{4\pi}\log D/r_0,\quad \text{(perturbative).}
\end{equation}
However this is cured by the fact that, near a delta function, there is a further logarithmic suppression of velocity, as discussed in Sec.~\ref{sec:log}. All in all, the resistance due to a sharp magnetic field distribution, for which $\omega_c\tau \log(D/r_0)\gtrsim 1$, is renormalized as 
\begin{equation}
f \to \frac{4\pi\log(D/r_0)}{(4\pi)^2 + (\omega_c\tau)^2\log^2(D/r_0)} ,\quad \text{(renormalized).}
\end{equation}

Once again, we shall take $r_0 \sim l_{ee}$ and therefore  $\log(D/r_0)\sim 1$. Working with logarithmic accuracy, we have essentially the same result as in Eq.~\eqref{fana}.
In order to have viscous resistivity to be comparable to the Drude resistivity, arising from non-magnetic impurities, we need, based on the Table~\ref{tab:params},
\begin{equation}
n_{m} \approx \frac{1}{4\pi D^2}\left(1+{\left(\frac{4\pi}{\omega_c\tau}\right)^2 }\right) \sim 10^{9}\:\text{cm}^{-2},\quad @ B \approx 20\:\text{mT},
\end{equation}
corresponding to about 10 micromagnetic impurities per micron squared. To compare, the electron number density is about a 200 times larger than this value. Realistically, if we increase the field to $B\approx 90\:\text{mT}$, we can achieve a $100\%$ correction to resistivity $\rho_{\nu,m}=\rho_\tau$ for 1 micromagnet per micron squared. 

\section{Conclusions}
	\label{sec:conc}
In this paper, we considered the response of the two-dimensional  viscous electron fluid to a localized out-of-plane magnetic field. From a purely theoretical perspective, this problem is akin to the Stokes problem for a 2D classical fluid. A drag force acts on the electron fluid arising from the scattering of electrons from the impurities and phonons in the system. The drag force acts on the long wavelength limit and cures the infrared divergence of the Stokes paradox.  The drag and viscous forces define a natural length scale, the momentum diffusion length $D$, in the system. We find that, when the magnetic field in concentrated in a region small compared to $D$, as in the case of a magnetic impurity, solutions perturbative in the field and gradient strength break down in the ultraviolet region. This is saved  by the logarithmic renormalization of fluid velocity near the impurity. Physically, this is because the large Lorentz force gradients near the impurity expels electric current. We demonstrate that this logarithmic suppression is crucial in predicting the experimental consequences of having micromagnets or magnetic adatom impurities in the system. We show that this effect introduces a large contribution to the resistivity of the system. Our findings are applicable to a wide range of platforms, including graphene and ultrapure semiconductor heterostructures.

\section*{ Acknowledgements} 
	
We acknowledge important discussions with Jack Engdahl. This work was supported by the Australian Research Council Centre of Excellence in Future Low-Energy Electronics Technologies (CE170100039). Multiphysics simulations are performed on the server of the New South Wales node of the Australian National Fabrication Facility.

\end{document}